\documentclass[final,5p,times,twocolumn,authoryear]{elsarticle}

\usepackage{amssymb}

\usepackage{graphicx}
\usepackage{txfonts}
\usepackage[colorlinks=true,linkcolor=blue,citecolor=blue, urlcolor=blue]{hyperref}
\usepackage[justification=centering]{caption}
\usepackage{physics}
\usepackage[separate-uncertainty=true, binary-units]{siunitx}
\usepackage{amsmath}
\usepackage{empheq}
\usepackage{enumitem}
\usepackage{subcaption}
\usepackage{xcolor}
\usepackage[ruled,vlined]{algorithm2e}
\usepackage{subcaption}
\usepackage{array}
\usepackage{float}
\usepackage{algorithm2e}
\usepackage{algpseudocode}
\usepackage{dblfloatfix}
\usepackage{placeins}

\usepackage{todonotes}

\newcommand\keto{\ket{0}}
\newcommand\keti{\ket{1}}

\journal{Astronomy $\&$ Computing}

\begin{document}

\begin{frontmatter}



\title{Quantum Radio Astronomy: Data Encodings and Quantum Image Processing}

\author[epfl]{Thomas Brunet}
\author[epfl,scitas]{Emma Tolley}
\author[scitas]{Stefano Corda}
\author[maastricht]{Roman Ilic}
\author[astron]{P. Chris Broekema}
\author[epfl]{Jean-Paul Kneib}

\affiliation[epfl]{organization={Institute of Physics, Laboratory of Astrophysics, École Polytechnique Fédérale de Lausanne (EPFL)},
            addressline={Observatoire de Sauverny}, 
            city={Versoix},
            postcode={1290}, 
            country={Switzerland}}

\affiliation[scitas]{organization={SCITAS, École Polytechnique Fédérale de Lausanne (EPFL)},
            city={Lausanne},
            postcode={1015}, 
            country={Switzerland}}
\affiliation[astron]{organization={Netherlands Institute for Radio Astronomy}, addressline={Oude Hoogeveensedijk 4}, 
            city={Dwingeloo},
            postcode={7991 PD}, 
            country={the Netherlands}}

\affiliation[maastricht]{organization={Department of Advanced Computing Sciences, Maastricht University}, addressline={Paul-Henri Spaaklaan 1}, 
            city={Maastricht},
            postcode={6229 EN}, 
            country={Netherlands}}



\begin{abstract}
We explore applications of quantum computing for radio interferometry and astronomy using recent developments in quantum image processing. We evaluate the suitability of different quantum image representations using a toy quantum computing image reconstruction pipeline, and compare its performance to the classical computing counterpart. For identifying and locating bright radio sources, quantum computing can offer an exponential speedup over classical algorithms, even when accounting for data encoding cost and repeated circuit evaluations.  We also propose a novel variational quantum computing algorithm for self-calibration of interferometer visibilities, and discuss future developments and research that would be necessary to make quantum computing for radio astronomy a reality.
\end{abstract}

\end{frontmatter}


\section{Introduction}

The exploitation of high-end, groundbreaking computing technologies has become essential for observational and theoretical astrophysics research. Next generation surveys and scientific instruments such as the  the Square Kilometer Array \citep{santander2017status},  the Cherenkov Telescope Array \citep{cta2011design},  and the James Webb Space Telescope \citep{gardner2006james} will produce exponentially more data than their predecessors, and will need outstanding resources to be post-processed, analyzed and stored. Moreover, numerical simulations, a theoretical counterpart capable of reproducing the formation and evolution of the astrophysical structures of our Universe~\citep{CosSim2020}, must be extended to larger volumes, higher resolutions and more sophisticated models to match the extensive data of the upcoming surveys.


 Quantum computing (QC) exploits principles of quantum mechanics to perform computational operations~\citep{QC1980,aharonov1999quantum}. In certain cases, algorithms designed to take advantage of QC can offer exponential speedup compared to their classical counterparts, known as {\em quantum advantage}.  Perhaps the most famous application of quantum computing is Shor's algorithm~\citep{shor1999polynomial}, which can solve prime number factorization in polynomial time. 
Recent years have witnessed rapid progress of quantum hardware technology, leading to the development of quantum processors that can support up to hundreds of qubits~\citep{chow2021ibm}.

 Exploiting QC solutions is a novel direction for Astrophysics. So far, quantum machine learning has been applied for the binary classification of pulsar data~\citep{qupulsar}, a hybrid quantum support vector machine has been proposed for classifying galactic morphology~\citep{qugal}, and an algorithm using variational quantum computing has been developed for dark matter cosmological simulations~\citep{qucos}. However, QC for radio interferometry remains completely unexplored.
 
 In this paper we develop and analyse QC data representations and algorithms in the context of radio interferometry, using recent developments in the field of quantum image processing (see \citep{wang2022review} for a recent review of this field). Our  goal is to evaluate the quantum advantage for radio astronomy, determining if QC algorithms and data encodings can ever achieve better performance than their classical counterparts. We also consider {\em quantum utility}, the effectiveness and practicality of QC algorithms on current or near-term hardware.

\section{Radio interferometry}

Most modern radio telescopes are radio interferometers, such as the Atacama Large Millimeter/submillimeter Array~(ALMA; \cite{2009alma, 2023alma}), 
the Very Large Array (VLA; \cite{selina2018next}), Low Frequency Array (LOFAR; \cite{van2013lofar}), MeerKAT \citep{jonas2016meerkat}, or the future Square Kilometer Array \citep{santander2017status}. 
Interferometry allows astronomers to bypass the inherent angular resolution of a single dish and obtain images of radio sources with very fine angular resolutions of up to milliarcseconds \citep{1958MNRAS}, allowing scientsits to image even the black hole at the center of our Galaxy~\citep{EHT_2022}. 
We refer readers to \cite{Thompson2017} for a more complete treatment of radio interferometry, and merely provide an abbreviated overview here.

Radio interferometers measure information about the sky in Fourier space, also called visibility space. The relation between the visibility space measurement and the brightness distribution of astrophysical radio sources is described by the van Cittert-Zernike theorem~\citep{Born1999}, which states that the two-point correlation function of the electric field measured by two antennas of a radio interferometer is the Fourier-transformed intensity distribution of the sources:
\begin{equation}
    V(u,v) = \int \hspace{-0.5em}\int S(x,y)~ e^{-2 \pi i ~(ux + vy)} ~dx~ dy
\end{equation}
where $S(x,y)$ is the image of the radio sky, $V(u,v)$ is the sampled visibility space, $x,y$  are angles in tangent plane relative to the pointing direction in the E-W and N-S directions, and $u,v$ are spatial frequencies in
E-W and N-S directions. We note that this is a simplification: the sky is a sphere, not a tangent plane.

By applying the inverse Fourier transformation to the data, one can reconstruct the image of the radio sky. 
First, the visibility measurements are resampled onto a regular grid $V \in  \mathbb{C}^{N\times N}$ using a gridding algorithm~\citep{cornwell1981,cornwell1992} such as IDG~\citep{idg2018}.
Then, the 2D discrete Fourier transform can be used to reconstruct the the matrix $S^D \in  \mathbb{R}^{N\times N}$:
\begin{equation}
    S^D_{lm} = \sum^{N-1}_{j=0} \sum^{N-1}_{k=0}  b_{jk} V_{jk}~ e^{- \frac{2 \pi i}{N} (jl + km)}
    \label{eq:ft}
\end{equation}
where $l,m$ and $j,k$ index the $x,y$ and $u,v$ coordinates, respectively, and $b_{jk} = \sum_\alpha \delta(j-j_\alpha,k-k_\alpha)$ is the visibility sampling function defined by the instrument geometry. The true sky model $S$ can be written as $S = B * S^D$, where $B$ is the point-spread function (or ``dirty beam'') of the instrument and defined as the Fourier transform of $b$. The true sky model $S^D$ must then be constructed by recursively deconvolving the dirty beam from the ``dirty image'' $S^D$ with a process like such as the CLEAN algorithm~\citep{clean1974}.

In the rest of the paper, we explore the advantages of replacing the 2D discrete Fourier transform of Equation~\ref{eq:ft} with a quantum algorithm.
Naively, reconstructing $S^D$ requires $N^4$ operations: each of the $N^2$ pixels in $S$ requires the sum of $N^2$ elements. However, for a regular grid this
 can be calculated with the fast Fourier transform algorithm in $\mathcal{O}( N^2 \log_2 N^2)$ = $\mathcal{O}( N^2 \log_2 N)$ time~\citep{fft1967}.

\section{Quantum computing}

 In classical computing, a bit at a given address can only hold one of two values: 0 or 1. In QC, the basic unit of information is the qubit, a quantum mechanical system  with two orthonormal basis states $\keto$ and $\keti$. 
 A single qubit represented by wavefunction  $\ket{\Psi}$ can be written as a superposition of these computational  basis states:
 \begin{equation}
     \ket{\Psi} =   \alpha\keto + \beta\keti
     \label{eq:simpleq}
 \end{equation}
where $\alpha$ and $\beta$ are complex probability amplitudes. By the Born rule, the probability of measuring $\ket{\Psi}$ to be in state $\keto$ is $|\alpha|^2$, and the probability of measuring $\ket{\Psi}$ to be in state $\keti$ is $|\beta|^2$. 
Exactly what the $\keto$ and $\keti$ basis states represent depends on the QC hardware. For example, a qubit could be represented by atomic spin, with $\keto$ representing the spin-down state and  $\keti$ representing the spin-up state~\citep{nmrqc,spinqc}.

Quantum states can be modified using unitary matrix operations called ``gates'' applied to single or multiple qubits. Some commonly used single qubit gates are the Hadamard gate $H$ ~\citep{hadamard2003} and the $R_x(\theta)/R_y(\theta)/R_z(\theta)$ rotation operator gates~\citep{rotgate1995}, and examples of multi-qubit gates are the $CNOT$ and 
$SWAP$ gates.
A quantum circuit is composed of an ordered sequence of these quantum gates. For example, the quantum state $\ket{\Psi}$ from Equation~\ref{eq:simpleq} can
be prepared from a $ \ket{\Psi} = \keto$  state by applying a Hadamard gate to create an equal superposition state $\ket{\Psi} =\frac{1}{\sqrt{2}}\keto + \frac{1}{\sqrt{2}}\keti$, and finally a rotation operator gate to set the correct relative amplitudes.
Encoding classical data into a multi-qubit quantum state is non-trivial and cannot always be performed efficiently. The time complexity depends on the encoding algorithm and on the data itself \citep{patterns}. 

Due to the randomness inherent in the quantum measurement, quantum states must often be prepared and measured multiple times, or ``shots,'' to obtain statistically robust results. For example, suppose one wants to measure the value of $|\alpha|^2$ from the quantum state $\ket{\Psi}$ in Equation~\ref{eq:simpleq}. A single measurement will collapse $\ket{\Psi}$ into $\keto$ or $\keti$, and the state $\ket{\Psi}$ must be prepared again to repeat the experiment. The value of $|\alpha|^2$ can then be estimated by comparing the number of times $\keto$ is measured compared to the total number of experiments, following binomial distribution statistics.

Real quantum hardware is noisy and prone to error. This means that in addition to the  uncertainty inherent in the measurement of a quantum state,  quantum gate errors can corrupt qubits of the quantum circuit. Typically the different quantum gates of a quantum computer will have different error rates, which much be taken into account to evaluate the realistic accuracy of a quantum algorithm.

\subsection{Quantum data representations}

Consider a classical image with $N^2$ pixels, which we represent as a 2D matrix $V_{ij}$, where $i$ and $j$ are the row and column indices, or as a flattened vector $V_k$, where $k$ is an index over all pixels. We define the normalized pixel values $c_k = V_k/\sqrt{\sum V_k^2}$.
A classical computer can represent the real-valued $V_k$ as an array of 64-bit floats, where each position in the array corresponds to a pixel in the image, and each value in the array represents the pixel value. The representation of an $N\times N$-pixel image requires $64N^2$ bits.

Classical data can be directly represented in a quantum state using {\bf binary encoding} \citep{loading}. Each bit with value 0 or 1 of the binary representations of the pixel values $V_k$ is simply represented by a corresponding quantum state $\keto$ or $\keti$, respectively.
In the case where $V_k$ are $64$-bit floats representing complex-valued visibility data, the entire image matrix can be encoded with $64N^2$ single-state qubits. 

Further data compression can be achieved with the {\bf quantum lattice} implementation~\citep{venegas2003storing}, in which the value of a pixel $c_k$ is represented by a single qubit:
\begin{equation}
 \ket{\Psi_k} = \cos{\theta_{k}}\keto + \sin{\theta_{k}}\keti
\end{equation}
where $\theta_{k} = \frac{\pi}{2} c_k$.
Thus the entire image can be encoded with $N^2$ qubits each in a superposition of two states. The measurement of the state returns either $\keto$ or $\keti$ with probability proportional to $\theta_{k}$. This state can easily be encoded using a Hadamard gate and a rotation operator gate as described above.

One of the most important aspects of quantum systems is a property known as entanglement. Entanglement between states is necessary to achieve any
significant computational speedup of QC over classical computing~\citep{loading,entaglement2003}.
Entanglement is used for the {\bf Flexible Representation of Quantum Images}  (FRQI; \cite{le2011flexible}) representation, encoding the positional information with $\log_2(N^2) + 1$ entangled qubits. The full image is represented by:
\vspace{-1em}
\begin{equation}
    \ket{\Psi} = \frac{1}{2^n}\sum_{k=0}^{N^2-1}(\cos{\theta_k}\keto + \sin{\theta_k}\keti) \otimes \ket{k}
    \label{eq:frqi}
\end{equation}
Where $\theta_k\in [0,\pi/2]$ is the scaled color information as before and $k=0,...,N^2-1$ indexes the computational basis vectors $\ket{k}$ which representing the pixel coordinates in binary strings of $\log_2(N^2)$ qubits.
For example, pixel 2 is represented by $\ket{k=2} =  \ket{0}\otimes...\otimes\ket{1}\otimes\ket{0}$. This encoding requires only $\log_2N^2 + 1$  qubits in a superposition of $2N^2$ states. Preparation of this state can be expensive, with a proposed circuit depth of $\mathcal{O}(N^4)$ operations in \cite{le2011flexible}, but can be reduced to $\mathcal{O}(N^2\mathrm{log}_2 N)$ operations by introducing additional ancillary qubits~\citep{frqi2019}.

\begin{figure}[h!]
    \centering
    \includegraphics[width=5cm]{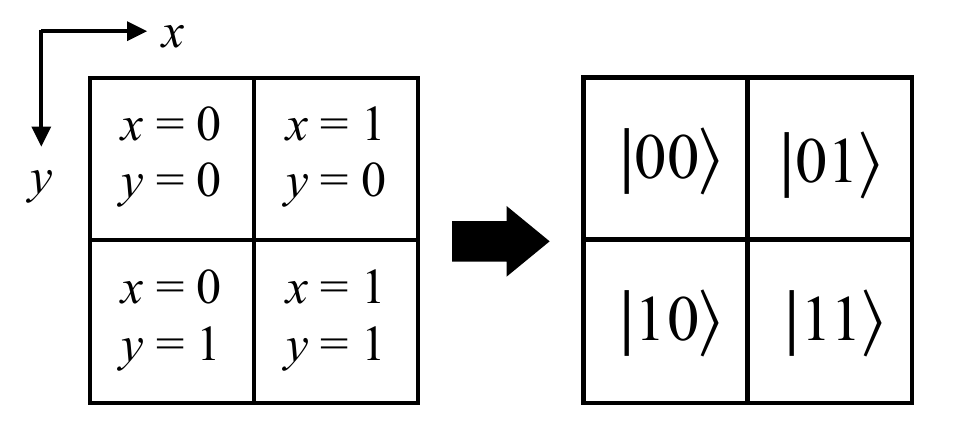}
    \caption{Scheme showing the separation of row and column indices.}
    \label{fig:schema_qft}
\end{figure}

The efficiency of quantum encoding is taken to its extreme with
{\bf Quantum Probability Image Encoding} (QPIE; \cite{yao2017quantum}). We encode an $N\times N$ image as an $N^2$-state superposition of $\log_2N^2$ qubits:
\vspace{-1em}
\begin{equation}
    \ket{\Psi} = \sum_{k=0}^{N^2-1}c_k\ket{k}
    \label{eq:qpie}
\end{equation}
where the amplitude of each state is determined by the corresponding pixel value $c_k = V_k/\sqrt{\sum V_k^2}$, for the flattened image vector $V$. By the Born rule, each measurement of the state will return a combination of qubits corresponding to the binary image index $i$, with probability proportional to the pixel value $V_i$. 
If $N$ is an even power of two, then we can rewrite the QPIE encoding to explicitly express the row and column indices:
\begin{equation}
    \ket{\Psi} = \sum_{x=0}^{N-1}\sum_{y=0}^{N-1}c_{i,j}\ket{i}\ket{j}
\label{eq:qpie2d}
\end{equation}
See Fig.(\ref{fig:schema_qft}) for an example of this coordinate mapping for a $2 \times 2$ image.

\noindent
Several different algorithms exist for preparing the QPIE representation, with different assumptions and tradeoffs for ancillary qubits, accuracy of the representation, and other parameters~\citep{grover2002creating,qpie2004,Shende_2006}.
Several state preparation algorithms with circuit depths varying from $\mathcal{O}( \log_2 N \times \log_2 N)$ to $\mathcal{O}( N^4)$ are proposed in \cite{stateprep2021}.

 There are many other quantum encodings for image representations (see \cite{wang2022review} and \cite{anand2022quantum} for reviews), but we will only consider the above quantum representations in the rest of this paper.

\subsection{Quantum Fourier transform}
The Quantum Fourier Transform (QFT) is the quantum analogue of the DFT. It is a quantum circuit which acts on the quantum state $\ket{X}=\sum_{j=0}^{M-1}x_j\ket{j}$ and maps it to the quantum state $\ket{Y}=\sum_{k=0}^{M-1}y_k\ket{k}$~\citep{nielsen2010quantum}, where the coefficients are given by:
\begin{equation*}
    y_k = \frac{1}{\sqrt{M}}\sum_{j=0}^{M-1}x_j ~ e^{2\pi i jk/M}
\end{equation*}
The QFT circuit requires $\mathcal{O}(n^2)$ gates, where $n$ is the number of qubits in the input and output states.

To apply the 2D QFT to the QPIE encoding of a 2D $N \times N$ 
image of Eq.~\ref{eq:qpie}, we can first apply the QFT to the row index qubits and then apply the QFT the column index qubits~\citep{qft2018}.
This is equivalent to applying a QFT to first half and second half of the qubits separately. 
As a QPIE encoding of an image with $N^2$ pixels only needs $\log_2 N^2$ qubits,  the QFT can be evaluated on this encoding with only $\mathcal{O}( \log_2 N \times \log_2 N)$ operations.


An implementation of the quantum FFT algorithm exists which has lower computational complexity with respect to the QFT \citep{qfft2020}, but it relies on the image being embedded in a basis encoding rather than the amplitude encoding used by QPIE.

\section{Experimental demonstrations}

Current QC hardware is often noisy and can only support small numbers of qubits~\citep{nisq2018}. On the IBM platform, the public can only access up to seven qubits, limiting us to representations with only $4 \times 4$ pixels for the QPIE encoding. Instead we run our experiments using the QASM simulator \citep{QSAM}.
This simulator emulates the execution of a quantum circuit on a real device. We do not include any modeling of gate or measurement errors. The code used for the demonstrations in Section~\ref{sec:tests} is available on GitHub\footnote{\href{https://github.com/QuantumRadioAstronomy/MA2-project_QuantumRadioImage}{https://github.com/QuantumRadioAstronomy/MA2-project\_QuantumRadioImage}}.

\subsection{Encoding \& decoding}
\label{sec:tests}

\begin{figure}[t]
    \centering
    \includegraphics[width=8.25cm]{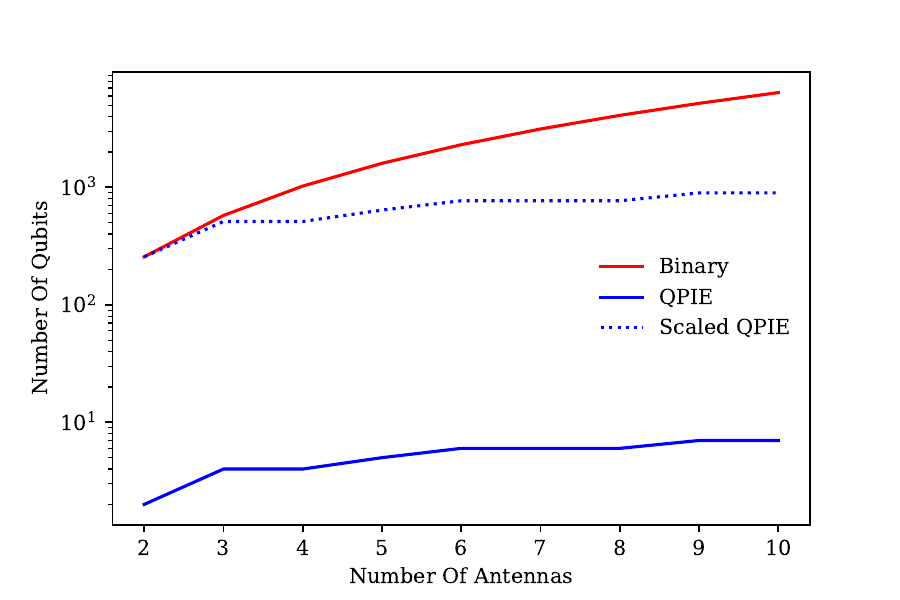}
    \caption{Evolution of the number of qubits required for QPIE and binary encoding for storing $M(M-1)/2$ correlated visibilities and their complex conjugates for increasing number of antennas $M$.  QPIE requires fewer qubits compared to the binary encoding, scaling logarithmically instead of quadratically with $M$. We also show the QPIE qubits scaled by a factor of 128 (dotted blue line) to better show the quadratic versus logarithic dependence.}
    \label{fig:schema_plot2}
\end{figure}

An incredible level of information compression can be achieved using quantum data encoding. For example, an interferometer with $M$ antennas will produce $B = M(M-1)/2$ correlated visibilities. In classical computing, the data for a given timestep can be encoded with $B$ $64$-bit floats representing complex-valued visibility data. However, the QPIE encoding leverages superposition to achieve logarithmic compression, only needing $\log_2(B)$ qubits, as shown in Fig.~\ref{fig:schema_plot2}.

 Encoding complex-valued numbers as complex probability amplitudes of a quantum superposition is straightforward, but the measurement of a quantum state can only return the real magnitude of the phases. In terms of practical applications of the QFT and encoding visibility data, this means that complex phases cannot be directly measured. For radio interferometry, while we can use QC to reconstruct real-valued images from complex visibilities (``invert'' step in data reduction), predicting complex visibilities from images (``predict'' step in data reduction) is not straightforward.

 Even if the QC can offer impressive data compression, measurement of a quantum state introduces inherent randomness. Quantum image representations must be re-encoded and re-measured multiple times to measure the state with high accuracy. This additional measurement error is a fundamental consequence of using QC compared to classical computing.
 
\begin{figure}[h]
    \centering
    \includegraphics[width=9cm]{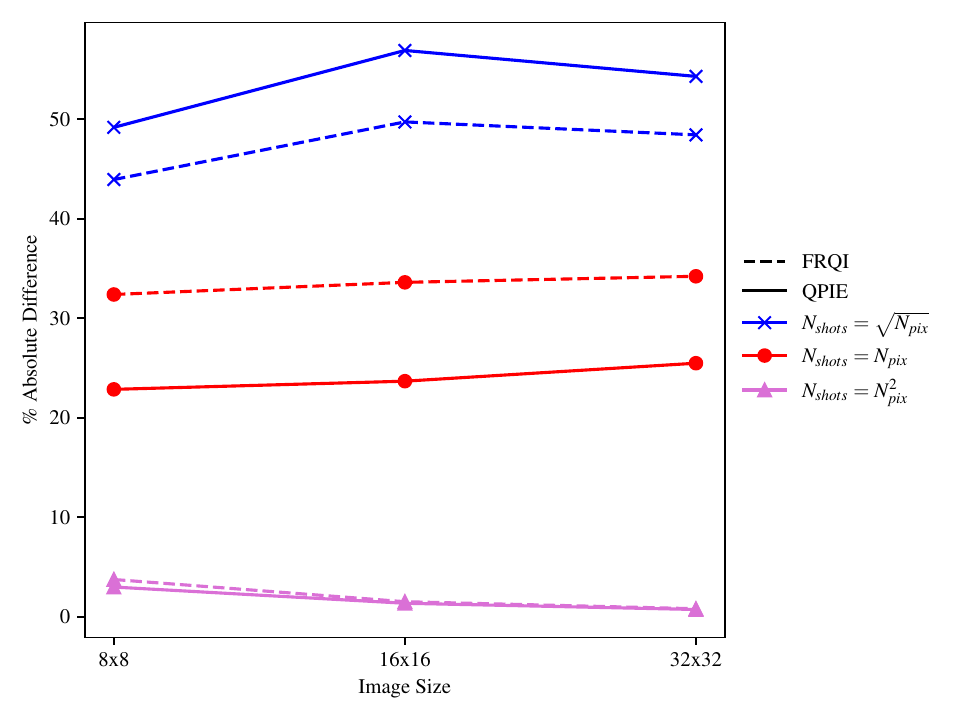}
    \caption{Image reconstruction accuracy as a function of $N_\mathrm{shots}$ and image size for the FRQI and QPIE encodings. Both encodings offer comparable accuracy, and require large numbers of shots  $N_{\mathrm{shots}} = N_\mathrm{pix}^2 = N^4$ to reconstruct a random input image with $< 10\%$ error. }
    \label{fig:acc_encodings}
\end{figure}

We  evaluate the number of measurements $N_{\mathrm{shots}}$ to create an accurate reconstruction of an image starting from the FRQI and QPIE encodings. We generate images of different sizes, with pixel values drawn from the uniform distribution $U(0,1)$, and compare the original with the quantum circuit output, shown in Fig.~\ref{fig:acc_encodings}. We see that both encodings offer comparable accuracy as a function of shot number. For both encodings, very large numbers of shots $N_{\mathrm{shots}} = N_\mathrm{pix}^2 = N^4$ are required to reconstruct a random input image with $< 10\%$ error. 

\subsection{Quantum images and source identification}

\begin{figure*}[ht]
\centering
    \includegraphics[width=17cm]{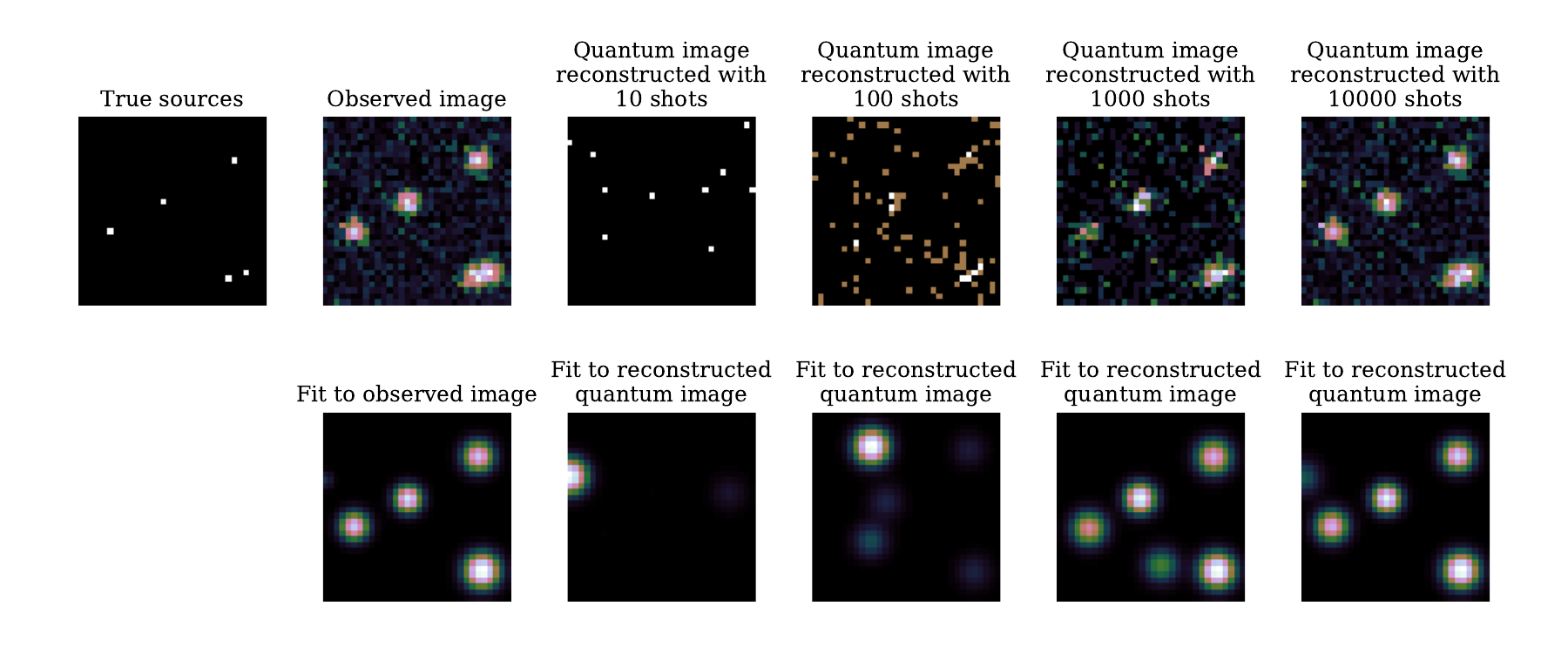}
    \caption{Mock source identification test with QPIE for a $N \times N = 32 \times 32$ pixel image. Five point sources are convolved with a Gaussian kernel with $\sigma=1.5$ pixels and Gaussian noise is added to create the mock observed image. This image is then embedded in a QPIE and reconstructed with varying numbers of shots to create the reconstructed quantum images. }
    \label{fig:obssim}
\end{figure*}
Astronomical images are often sparse and noisy, and reliable characterizations of radio sources do not necessarily need measurements of every single pixel in the image.
We developed a mock source fitting pipeline in the style of pyBDSF~\citep{mohan2015pybdsf}  or Agean~\citep{agean2018} to evaluate the performance and tradeoffs of quantum images, available on GitHub\footnote{\href{https://github.com/QuantumRadioAstronomy/QCRadioSimulator}{https://github.com/QuantumRadioAstronomy/QCRadioSimulator}}. Randomly placed point sources are convolved with a Gaussian kernel with $\sigma=1.5$ pixels. Then, random Gaussian noise is added to create a mock observed image. This image is then embedded in a QPIE, FRQI, or Quantum Lattice representation and reconstructed with varying numbers of shots to create reconstructed quantum images. We recursively fit 2D Gaussians to the reconstructed images to identify all of the sources. Example images from this simple simulation are shown in Fig.~\ref{fig:obssim}.

\begin{figure*}[h!]
\centering
    \begin{subfigure}{8cm}
    \includegraphics[width=8cm]{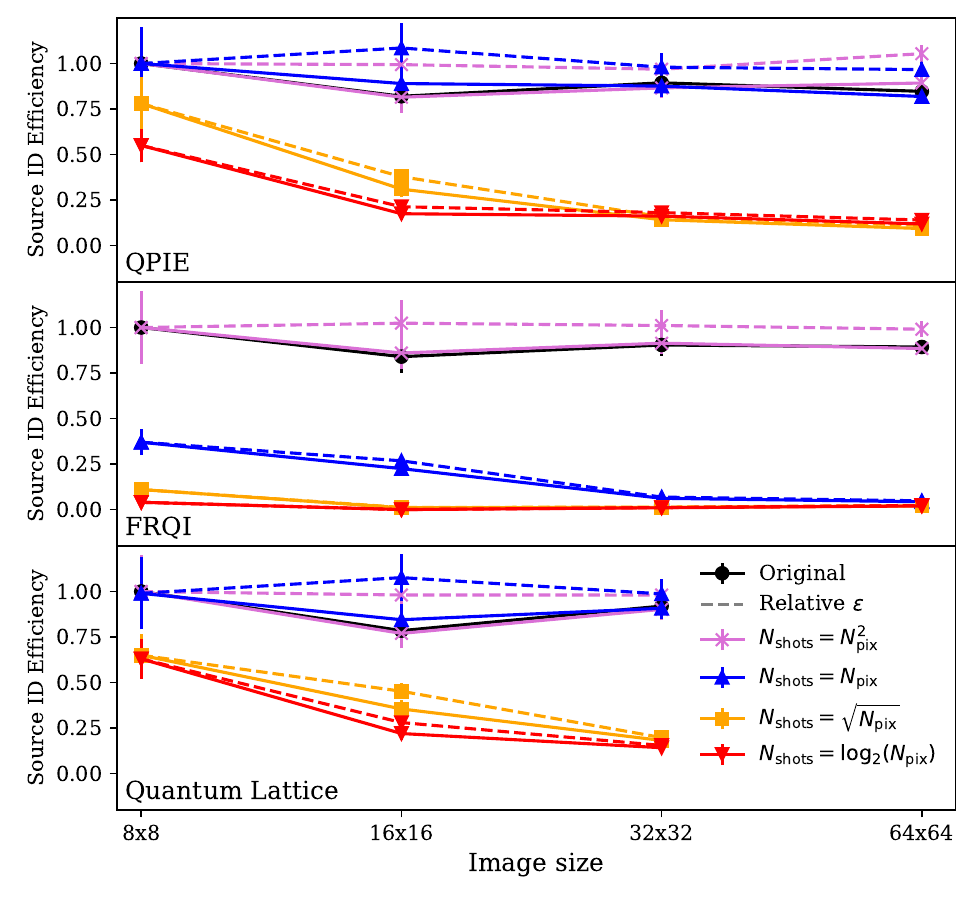}
    \caption{Source reconstruction efficiency for $N_\mathrm{source} \propto N^2$ sources with SNR = 10.}
    \label{fig:obssimsizeA}
    \end{subfigure}
\begin{subfigure}{8cm}
    \includegraphics[width=8cm]{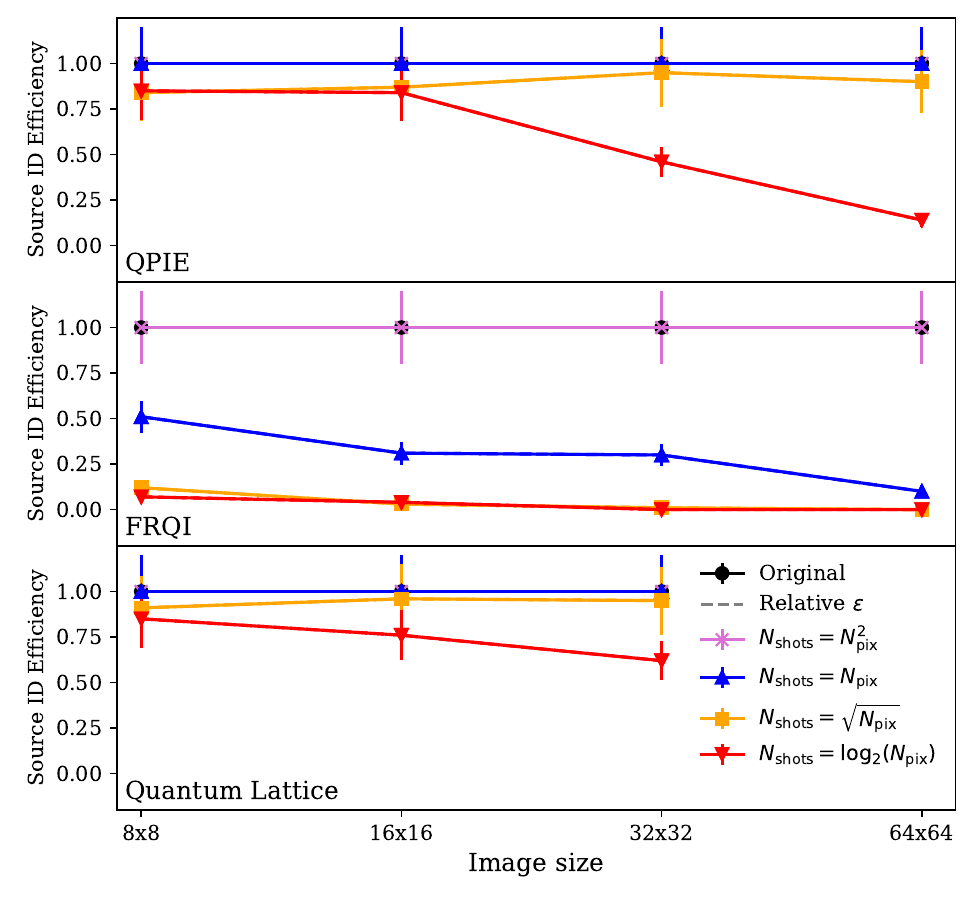}
    \caption{Source reconstruction efficiency for one $N_\mathrm{source} = 1$ with SNR = 100.}
    \label{fig:obssimsizeB}
    \end{subfigure}
    \caption{Efficiency of source reconstruction for classical vs quantum computing as a function of image size $N_\mathrm{pix} = N^2$ for three different quantum image encodings: QPIE, FRQI, and Quantum Lattice. For both source identification scenarios, the number of shots used to reconstruct the quantum image is scaled as a function of image size in order to evaluate the scaling relation. A source is considered correctly identified if it is reconstructed with a 1.5 pixel distance of the true source. The dotted lines reflect the relative efficiency of the quantum implementation with respect to the classical implementation. For the noisy SNR=10 images occasionally we observe that the quantum implementation has better performance than the classical image. Occasionally overlapping sources will not be recovered by the classical algorithm but will be recovered by a low-shot quantum algorithm.   }
    \label{fig:obssimsize}
\end{figure*}

The distance between the reconstructed and true sources are used as a metric of image quality. We consider that a source is correctly identified if it is reconstructed with a 1.5 pixel distance from the true source.
%
%
We evaluate the $N_\mathrm{shot}$ scaling with image size vs efficiency in Fig.~\ref{fig:obssimsize} for two source finding scenarios. In the first scenario, we scale the number of sources with image size $N_\mathrm{source} \propto  N_\mathrm{pix} = N^2$ and have a lower SNR=10 for each source, shown in Fig.~\ref{fig:obssimsizeA}. This model approximates a typical radio source catalog creation pipeline.
In the second scenario, we have a single source with a high SNR=100, shown in Fig.~\ref{fig:obssimsizeB}. This model approximates searches for bright transient phenomena in the image domain.

For QPIE, we see that when $N_\mathrm{shot} = N_\mathrm{pixel} = N^2$, the reconstructed quantum image is of good quality with respect to the classical image. In images with high signal-to-noise, even fewer shots can be taken. We see comparable source identification efficiency for $N_\mathrm{shot} =  N^2$ and $N_\mathrm{shot} =  N^2/2$ for low-noise images. As the noise increases, $N_\mathrm{shot} =  N^2$ retains good efficiency, but the other sampling scenarios suffer in performance.

The FRQI representation has a high probability to return pixels associated with low flux values, and therefore many shots are required to reconstruct the sources. We find that the FRQI representation needs at least $N_\mathrm{shot} =  N_\mathrm{pix}^2 = N^4$ shots.

The Quantum Lattice representation shows comparable sampling scaling to the QPIE representation, despite the drastically different number of qubits used by each encoding.

Our model pipeline is a simple approximation of real scientific use case of identifying radio sources in large-scale surveys or searching for transients in the image plane~\citep{lofartransient2015}.
We find that for these sparse images that the Quantum Lattice and QPIE quantum data representations can offer comparable source reconstruction performance with respect to classical computing without requiring an intractable number of shots.  However, measurements of extended structure or cosmological signals will require high-dynamic range, high-fidelity images which may require $N_\text{shots} > N_\text{pix}^2$ and thus are not suited for the quantum image encodings discussed in this work.


\begin{table*}[h!]
\smaller
\begin{center}
\begin{tabular}{|c|c|c|c|c|c|}\hline 
  {\bf Image Encoding } & {\bf Quantum Lattice} & \multicolumn{2}{|c|}{\bf FRQI} & \multicolumn{2}{|c|}{\bf QPIE}\\
  \hline
\# of qubits for representation & $N^2$ & \multicolumn{2}{|c|}{$\log_2N^2 + 1$} & \multicolumn{2}{|c|}{$\log_2N^2$}\\
\hline
Encoding circuit depth
    & 1
    & $\mathcal{O}(N^4)$ & $\mathcal{O}(N^2\mathrm{log}_2 N)$ 
    &  $\mathcal{O}(N^4)$ 
    &
    $\mathcal{O}(\mathrm{log}_2 N \times \mathrm{log}_2 N)$   \\
    \hline
Ancillary qubits for encoding circuit & None & None & $\mathcal{O}(\mathrm{log}_2N)$ & $\mathcal{O}(\mathrm{log}_2N)$ &$\mathcal{O}(N^4)$ \\

\hline
$N_\mathrm{shots}$ for multi-source reconstruction & $\mathcal{O}(N^2) $ & \multicolumn{2}{|c|}{$\mathcal{O}(N^4)$} & \multicolumn{2}{|c|}{$\mathcal{O}(N^2)$} \\
\hline
$N_\mathrm{shots}$ for single bright source & $\mathcal{O}(N)$ & \multicolumn{2}{|c|}{$\mathcal{O}(N^4)$} & \multicolumn{2}{|c|}{$\mathcal{O}(N)$}\\
\hline
\end{tabular}
\end{center}
\caption{The computational complexity of different stages of quantum computation represented in Fig.~\ref{fig:qa} and studied in this work. Encoding the FRQI state requires a circuit depth of $\mathcal{O}(2^{4n}) = \mathcal{O}(N^4)$~\citep{le2011flexible}, but can be reduced to $\mathcal{O}(N^2\mathrm{log}_2 N)$ operations by introducing additional ancillary qubits~\citep{frqi2019}. There are many algorithms for QPIE state preparation, but we pick two with circuit depths varying from $\mathcal{O}( N^4)$ to $\mathcal{O}( \log_2 N \times \log_2 N)$ from \citep{stateprep2021}.}
\label{tab:results3}
\end{table*}

\section{Quantum Advantage}
\begin{figure}[h!]
    \centering
    \includegraphics[width=6cm]{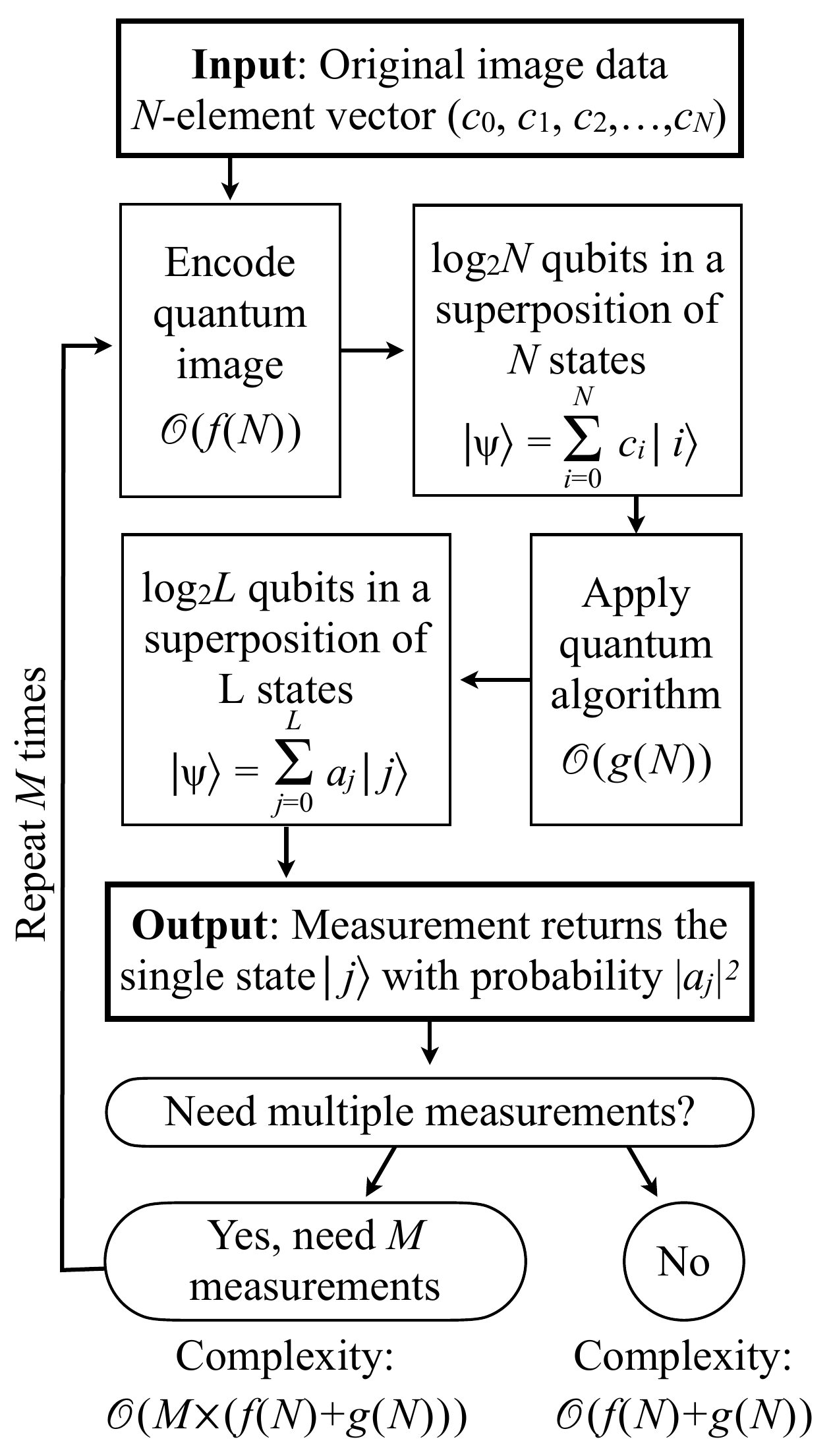}
    \caption{Scheme showing the computational complexity of encoding and measurements.}
    \label{fig:qa}
\end{figure}

Evaluating quantum advantage is not as simple as comparing the algorithmic complexity between a classical and quantum algorithm. Information must also be extracted from the quantum state, and measurement destroys the superposition. To generate multiple measurements of a quantum image, the entire image representation must be re-encoded and re-measured. Thus the cost of the state preparation and the number of shots needed to reconstruct the image must be taken account into the algorithmic complexity, as shown in Fig.~\ref{fig:qa}.
We note that it is possible that this encoding cost may be bypassed with quantum random access memory, allowing the input data can be stored in ancillary qubits and copied instead of expensive re-encoding. However, we consider our model in Fig.~\ref{fig:qa} realistic for near-term quantum computers.


A summary of the algorithmic complexities for the different quantum image encoding strategies is shown in Table~\ref{tab:results3}.
Let us consider the case of reconstructing the real-valued dirty image $S^D$ from the complex-valued measured visibilities $V$. In the classical domain, this operation requires $\mathcal{O}(N^4)$ operations using the DFT, or $\mathcal{O}(N^2 \mathrm{log}_2 N)$ operations using the FFT.

The QFT offers impressive speedup, only needing $\mathcal{O}(g(N^2)) = \mathcal{O}(\mathrm{log}_2 N \times \mathrm{log}_2 N)$ operations on the QPIE encoding, outperforming both the DFT and the FFT. However, the cost of the state preparation and the number of shots needed to reconstruct the image must be taken account into the algorithmic complexity, as shown in Fig.~\ref{fig:qa}. If we consider the optimal QPIE circuit depth of $\mathcal{O}(f(N^2)) = \mathcal{O}(\mathrm{log}_2 N \times \mathrm{log}_2 N)$, and the optimal number of shots needed to reconstruct sources $N_\mathrm{shots}$ the true complexity of the QFT is:
\begin{equation}
   \mathcal{O}( M \times (f(N^2) + g(N^2))) = \mathcal{O}(N_\mathrm{shots} \times \mathrm{log}_2 N \times \mathrm{log}_2 N)
\end{equation}
Depending on the number of shots needed to perform the measurement, the QFT may not offer quantum advantage. If a high-fidelity image needs to be extracted from the quantum computer, then $N_\mathrm{shots}$ becomes intractably large. However, if only a few samples of the image are needed, the problem becomes more manageable. In the case of identifying $\propto N^2$ sources in an image, $N_\mathrm{shots} = \mathcal{O}(N^2)$ shots are needed, which has quantum advantage over the DFT but not the FFT.  In the case of identifying one bright source in an image, $N_\mathrm{shots} = \mathcal{O}(N)$ shots are needed, offering exponential speedup over the FFT.

The number of shots required will not change as quantum hardware improves, but can be improved with algorithmic development. 
It is advantageous to stay in the quantum domain for as long as possible before measuring the state, ideally performing additional data reduction so that fewer shots are required for analysis. This could include performing gridding, deconvolving $S^D$ to reconstruct $S$, running additional image analysis algorithms, doing quantum machine learning, etc.

\section{Quantum gain calibration}
\label{sec:qselfcal}

Reconstructing an image of the radio sky is more complex than simply applying the DFT to the antenna visibilities.
The electric field measured by a single radio antenna is not constant with time. Drifting antenna-specific gain terms $g_i$ can distort the true signal, and need to be corrected through calibration. As a proof-of-concept, we
 define and implement a quantum algorithm for calibration using quantum variational circuits \citep{var_circ}. 

\subsection{Classical Calibration}
If each antenna $i$ in a radio interferometer measures a voltage $v_i$, then the measured visibilities are correlations between those voltages $V_{jk} = E[v_j v_k^*]$. When antenna-specific gain terms distort these measurements, the observed visibilities will be $\widetilde V_{jk} = g_j g_k^* V_{jk}$~\citep{smirnov2011}. 

We correct for the unknown gain terms by observing a known source $I_{pq}$ with the instrument, and calculating the expected visibilities with the DFT: $V_{jk} = \mathrm{DFT} (I_{pq})$.
The complex gains can be determined by minimizing the residual of the least-squares optimization problem:
\begin{equation}
    R = \sum_{j,k}\bigl|\widetilde{V}_{jk}-g_jg_k^*V_{jk}\bigr|^2
    \label{eq:lsq}
\end{equation}
$\widetilde{V}$ are the known, observed visibilities, $V$ are the expected true visibilities given by $V=\mathrm{DFT}(I)$, and $g_j,g_k$ are the unknown gain factors. 

In this toy model, we will assume that the interferometer geometry samples every point in Fourier space, allowing us to represent $V$ using the 2D image QPIE encoding. However, this technique can easily be extended to more realistic interferometer models with a different choice of quantum data encoding.

\subsection{Quantum Implementation}

\begin{figure}[h!]
    \centering
    \includegraphics[width=5cm]{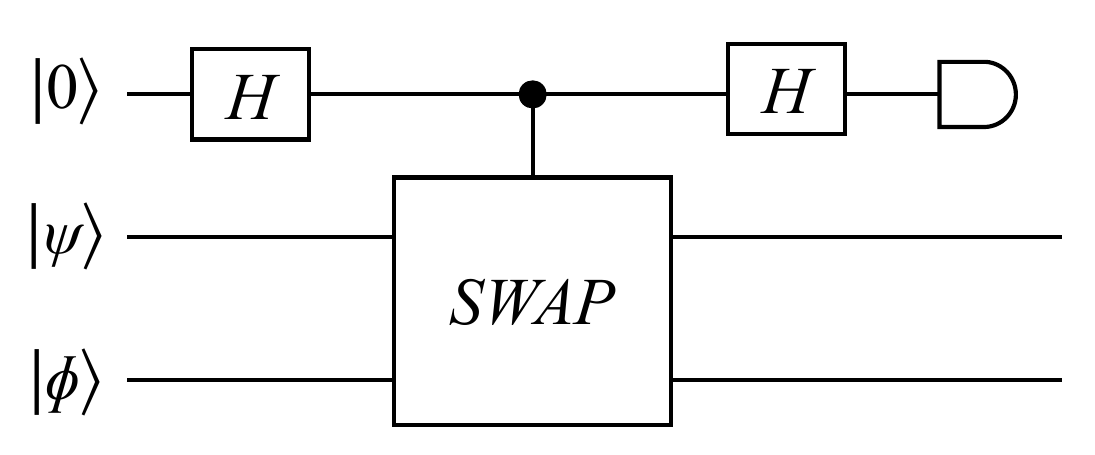}
    \caption{The SWAP Test circuit.}
    \label{fig:swap:circ}
\end{figure}

We define our variational circuit using the QPIE encodings corresponding to $\ket{\psi}=\widetilde V_{ij}$ and $\ket{\phi}=g_ig_j^*V_{ij}$, combined with the SWAP test~\citep{buhrman2001quantum}.
The SWAP Test allows one to compute the quantity $|\bra{\psi}\ket{\phi}|^2$ between two quantum states, and its circuit is shown in Fig.~\ref{fig:swap:circ}.

In this circuit, the measurement of the first qubit will yield $\keti$  with probability  $P(\keti)=\frac{1}{2}-\frac{1}{2}|\bra{\psi}\ket{\phi}|^2$ \citep{buhrman2001quantum}. We define the cost function to be minimized as follows:
\begin{equation}
    C(\{g_i\}) = P(\keti)=\frac{1}{2}-\frac{1}{2}|\bra{\psi}\ket{\phi}|^2
    \label{eq:swaptest}
\end{equation}
Where $\ket{\psi}$ and $\ket{\phi}$ are the QPIE states of $g_ig_j^*V_{ij}$ and $\widetilde{V}_{ij}$, respectively.  One can find the parameters $g_i$ so that $\widetilde{V}_{ij}$ and $g_ig_j^*V_{ij}$ are as close as possible by maximizing the dot product $|\bra{\psi}\ket{\phi}|^2$, and thus minimizing $ C(\{g_i\}) $.

We minimize this cost function with quantum gradient descent~(QGD; \cite{QGD}).
Given the measurement of our variational circuit $ C(\{g_i\}) $ at step $k$, the parameters are updated at step $k+1$ with the rule $g^{(k+1)}_{i} = g^k_{i} - \eta \,\partial_{g_i}C(\{g_i\})$, until convergence, where $\eta$ is the learning rate parameter. In the quantum implementation, the gradients are computed using parameter shift rule \citep{psr} given by:
\begin{equation}
\begin{split}
    \partial_{g_i}C(\{g_i\}) &=  \partial_{g_i}C(g_1,...,g_i,...,g_n) \\
    & = m[C(g_1,...,g_i+s,...,g_n)-C(g_1,...,g_i-s,...,g_n)]
\end{split}
\end{equation}
where the multiplier $m$ and the shift $s$ are typically chosen to be $m = 1/2$ and $s=\pi/2$. 
It is important to emphasize that in this approach, the shifts employed are not infinitesimal as typically used in finite differences. This technique facilitates the computation of gradients in a straightforward manner.
The QGD algorithm has a theoretical computational complexity of $\mathcal{O}(1)$~\citep{qgrad2005} for evaluating the gradient, which provides significant advantages over classical algorithm with a computational complexity of $\mathcal{O}(d)$, where $d$ is the number of input parameters.

The full algorithm is shown in Alg.~\ref{alg:one}. Because $C(\{g_i\})$ must be measured through repeated measurements of our variational circuit, the efficiency of this algorithm strongly depends on the data encoding strategy. 

\begin{algorithm}
\caption{Quantum Gain Calibration}\label{alg:one}
  \DontPrintSemicolon

\KwData{$\widetilde{V}$, $V$}
\KwResult{$\{g_i\}$ minimizing $\sum_{i,j}\bigl|\widetilde{V}_{ij}-g_ig_j^*V_{ij}\bigr|^2$}

  \SetKwFunction{FMain}{SWAPtest}
  \SetKwFunction{FTest}{CostFunc}
  \SetKwFunction{FShift}{ParamShift}
  \SetKwProg{Fn}{Function}{:}{}
  \Fn{\FMain{$\{g_i\}$}}{
        $\ket{\psi} \gets g_i g_j^* V_{ij}$ in QPIE encoding\;
        $\ket{\phi} \gets \widetilde{V}_{ij}$ in QPIE encoding\;
        Perform SWAP test on $\ket{\psi}$ and $\ket{\phi}$\;
        \KwRet measurement of first qubit with $P(\keti)=\frac{1}{2}-\frac{1}{2}|\bra{\psi}\ket{\phi}|^2$, $P(\keto) = 1-P(\keti)$\;
  }
    \Fn{\FTest{$\{g_i\}$}}{
        $C\gets 0$\;
        \For{$N$ iterations}{
         $C\gets C + $\FMain{$\{g_i\}$, $\widetilde{V}$, $V$}\;
        }
        \KwRet $C/N$\;
  }
      \Fn{\FShift{$\{g_i\}$}}{
        $G^+\gets (g_0,...,g_0 + s,...,g_n)$\;
        $G^-\gets (g_0,...,g_0 - s,...,g_n)$\;
        \KwRet $m \times (\FTest(G^+) - \FTest(G^-))$\;
  }
  \;

$\{g_i\} \gets$ random initial values\;
\While{$C(\{g_i\}) \geq \epsilon$}{
    $\{g_i\} \gets \{g_i\} + $\FShift{$\{g_i\}$}  \;
    $C(\{g_i\}) \gets$ \FTest($\{g_i\}$, $\widetilde{V}$, $V$)  \;
}
\end{algorithm}

\subsection{Hybrid Implementation}
 We also define a hybrid quantum-classical approach. This approach is very similar to the pure quantum method.
Given the observed visibilities $\widetilde{V}_{ij}$ the expected true visibilities $V_{ij}$, and an initial guess at the gain terms $\{g_i\}$, we encode $\ket{\psi}=\widetilde V_{ij}$ and $\ket{\phi}=g_ig_j^*V_{ij}$ using the QPIE encoding. Then, we
evaluate the cost function $C(\{g_i\})$ using the $SWAP$ test of Equation~\ref{eq:swaptest} $N$ times to accurately measure $C(\{g_i\})$. However, unlike the pure quantum algorithm, instead of using QGD to find the $\{g_i\}$ which minimize $C(\{g_i\})$, we use the COBYLA minimizer on Scipy.

Due to the quantum uncertainty inherent in the state measurements, neither calibration method is expected to offer quantum advantage over a classical minimization algorithm. The quantum variational circuit must be chained together with the circuit for another algorithm such as the QFT to obtain quantum advantage.

\subsection{Results}
We evaluate the performance of our quantum calibration algorithm using the  QASM simulator. We compare our pure quantum QGD algorithm to the hybrid quantum-classical approach. To simplify the analysis we assume that the gain terms ${g_i}$ are real numbers, but our algorithm can easily be generalized to complex gains by treating each complex gain as two parameters representing the real and imaginary parts.

\begin{figure}[h!]
    \centering
    \includegraphics[width=8cm]{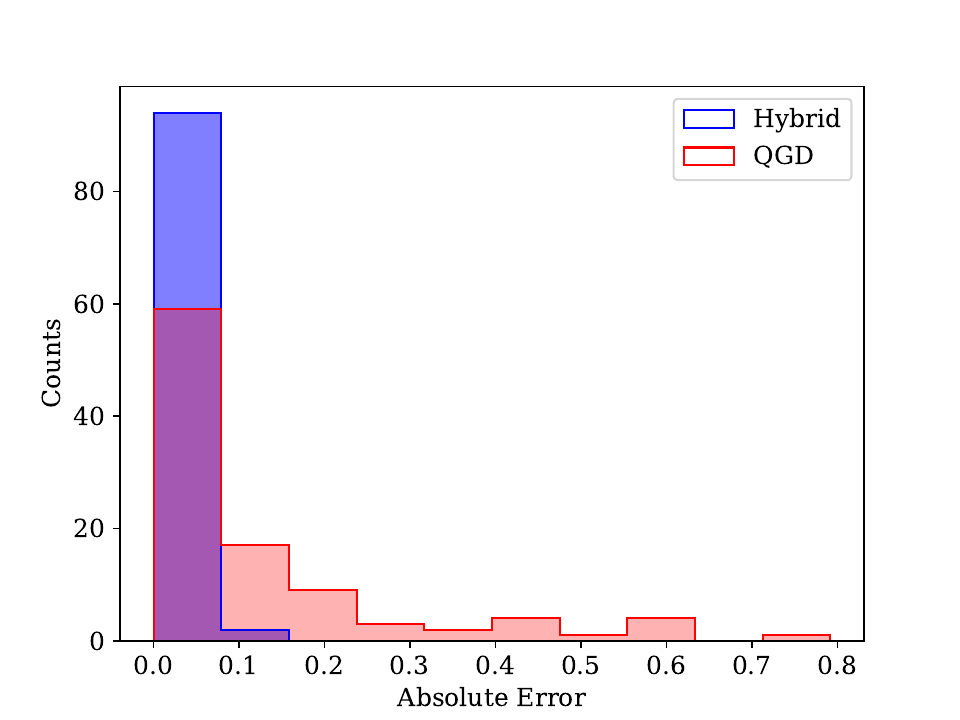}
    \caption{Error on recovered gain amplitudes upon the convergence of multiple trials of the pure QC and hybrid -calibration algorithms. While both approaches are able to recover the gain facros with high accuracy, the hybrid approach shows more stability.}
    \label{fig:selfcalres}
\end{figure}

We evaluate the performance using randomly generated $2\times 2$ radio sky images.
QGD is run for 1000 steps and with $\eta=0.01$. 
The error on the obtained gains for various random images are presented in Fig.~\ref{fig:selfcalres}.
The algorithm performs  well and can successfully recover the gain factors with high accuracy. 
 However, the runtime for QGD is quite long. Evaluating 1000 steps takes approximately two minutes on the quantum simulator. In comparison, the hybrid approach converges in approximately five seconds.

While the fully quantum algorithm employing QGD may appear inefficient due to the significant number of required iterations, the hybrid approach involving a classical optimizer demonstrates promising results. We note, however, that for this toy example using a purely classical approach is more efficient and accurate. However, this variational quantum circuit approach can be extended to include algorithms such as the QFT, possibly offering quantum advantage.

\section{Quantum Utility}
While quantum image processing offers considerable computational improvements for radio interferometry, the high error rates of real quantum systems limit quantum utility. Environmental effects can corrupt the quantum gates, leading to errors in the quantum circuit. In the previous sections we have ignored the effect of this quantum error.

Assuming that each gate has an error rate of $\epsilon = P(\text{gate fails})$ and a circuit depth of $D_\text{circ}$ the global error rate is approximately:
\begin{equation}
\begin{split}
    P(\text{at least 1 gate fails}) &= 1 - P(\text{all gates succeed}) \\
    & = 1- (1-\epsilon)^{D_\text{circ} }
\end{split}
\end{equation}
However, the noise for QC is more complex than statistically independent failures~\citep{morvan2023phase}, and for a complete analysis must be evaluated on real quantum hardware. The above approach also ignores the effect of measurement and decoherence errors, and thus should only be considered an approximate model of the true expected error rate.

We evaluate the approximate error rate for our QPIE encoding. We build the QPIE encoding quantum circuit using a recursive initialization algorithm from \cite{Shende_2006}. This circuit is constructed from $R_z(\theta)$, $SX$\footnote{The $SX$ gate is a special case of $R_x(\theta)$ Gate where the rotation parameter $\theta$ (the phase change) is 90° or $\pi/2$ radians.}, and $CNOT$ gates.
%
%
The gate errors of the 5 qubit IBM Falcon r4T system have been measured through randomized benchmarking~\citep{2012qerr} and are publically available\footnote{\href{https://quantum-computing.ibm.com/services/resources?tab=systems }{https://quantum-computing.ibm.com/services/resources?tab=systems }}: $CNOT$ has a median error rate of $9.286\times 10^{-3}$, and $R_z(\theta)$ and $SX$ of $4.175\times10^{-4}$. 
We calculate that encoding a $4\times 4$ image with the recursive initialization algorithm requires 30 $R_z(\theta)$, 30 $SX$ and 14 $CNOT$ gates. Using the error rates above, the probability that the circuit fails is $\approx 15\%$. 

If the gate error is incoherent and uncorrelated\footnote{While gate noise can involve coherent errors such as systematic gate bias, such errors are hardware-specific}, then then the effect of this error on the measurement can be reduced by taking additional shots~\citep{Sharma_2020}. 
It may be possible to mitigate the effect of quantum error by using quantum error correction techniques to identify corrupted circuits~\citep{qerrcorr}. However, these corrective techniques will not work if the circuit failure rate is too high.

A larger $256 \times 256$ image requires quantum hardware with at least 16 qubits, such as the IBM  Falcon r5.11 system. This system has published error rates of $2.091\times10^{-4}$ for the $SX$ and $R_z(\theta)$ and $8.698\times10^{-3}$ for the $CNOT$ gates. We calculate that encoding a QPIE image $256 \times 256$ requires a circuit with 131070 $R_z(\theta)$, 131070 $SX$, and 65534 $CNOT$ gates. With such a deep circuit, the probability that this circuit fails is $\approx 100\%$. The large circuit depth and high gate error rate make it impossible to encode the larger QPIE image.

This situation may be improved by using the shorter circuits proposed by~\cite{stateprep2021}, which provide  $D_\text{circ} = \mathcal{O}(\mathrm{log}_2N\times \mathrm{log}_2N)$ algorithms for QPIE encoding. With such a circuit, a $1024 \times 1024$ image can be encoded with a circuit depth of $\sim 100$ gates, and thus even with a gate error of $\epsilon=10^{-4}$ the global failure rate will be $\sim1\%$. However, the large number of ancillary qubits required by this algorithm make it impractical to run on current hardware, and may be subject to additional decoherence errors.

Thus the quantum utility of quantum image processing is limited to smaller image sizes by the noise and size of available hardware. However, because the Fourier transform is a linear operation, a single Fourier transform on a large image can be split into multiple operations on smaller image sizes~\citep{hvox2023,bipp2023}. 
It may be possible to perform equivalent input splitting with the QC, allowing hybrid QC algorithms to accommodate larger images. However, QC hardware will still need to be able to encode and process large input images in order to offer better performance compared to classical computing solutions.

\section{Conclusion}
Quantum Computing has promising applications for image processing in radio astronomy, but is limited by the error rates and small qubit sizes of contemporary noisy intermediate-scale quantum computers~\citep{nisq2018}. We have investigated the applicability of the QFT to image synthesis, implemented different image representations in a quantum computer, and evaluated their performance with respect to a simple source identification pipeline. Of the different quantum image encodings, we find that the QPIE encoding offers excellent algorithmic speedup without sacrificing image fidelity.

However, QC is not without its own complications. The phases of complex probability amplitudes can not be easily read from a quantum state, limiting the applications of the inverse QFT for predicting visibilities from images. 
Additionally, although the QFT offers exponential speedup over the FFT, measurement of the quantum state is a probabilistic result which destroys the superposition. To perform multiple measurements, one must re-encode and re-measure the state, which can undermine quantum advantage. It is possible that future developments in quantum algorithms and quantum hardware, especially quantum memory, may alleviate the re-encoding requirement. 

Despite these limitations, we have demonstrated that in cases where lower image accuracy is tolerable quantum computing can be a powerful tool for real-time imaging. In the case of identifying bright transients in the image domain, QC can offer an exponential speedup over classical computing. The QPIE encodings may not be suitable for high-fidelity image analysis demanded by extended source structure or cosmological studies, but other quantum information encodings may prove more suitable for these applications~\citep{schuld2018supervised}.

Additional algorithmic developments in the field of quantum computing could improve this situation further. Of particular interest for imaging in radio astronomy are:
\begin{itemize}
\item A quantum algorithm for gridding and/or degridding of the input visibilities
\item An implementation of the QFFT that can be applied to amplitude-based image encodings such as the QPIE.
\item A quantum implementation of the non-uniform discrete Fourier transform, allowing us to skip the the gridding step required to map the visibilities to a uniform grid. 
\item An efficient algorithm for QPIE encoding of sparse images
\item A quantum algorithm for deconvolution of $S^D$ with the dirty beam
\end{itemize}
Additionally, a wide range of developments in Quantum Machine Learning \citep{schuld2015introduction} could be useful for image analysis.

While the practical quantum utility of these algorithms is limited by the small qubit size and noise of contemporary quantum computers, QC technology is a rapidly developing field. As systems are released with more qubits and less noise, QC could provide a solution to the future data processing challenges of radio astronomy.

\section*{Acknowledgements}
This work has been done in partnership with the SKACH consortium through funding by SERI, and was supported by EPFL through the use of the facilities of its Scientific IT and Application Support Center (SCITAS). RI would also like to thank Dr. Daniel C\'ampora P\'erez (University of Maastricht)  for helpful discussions and suggestions.  The authors would also like to
thank the referee for their valuable comments and suggestions that
have improved the quality of this manuscript.

This paper is part of a broader investigation into the feasiblity of quantum computers in radio astronomy. We thank Dr. Nicolas Renaud, Dr. Pablo Rodr\'iguez-S\'anchez, and Dr. Johan Hidding (Netherlands eScience Center) for the interesting discussions, and refer to their
accompanying paper in this same journal.

\bibliographystyle{elsarticle-harv} 
\bibliography{references}

\end{document}